# Two Cases of Radial Adiabatic Motions of a Polytrope with Gamma=4/3

Mikhail I. Ivanov[*]

**ABSTRACT**

A self-gravitating sphere of polytropic gas (polytrope) is considered. The system of equations describing radial motions of this sphere in Lagrangian variables reduces to the only nonlinear PDE of the second order in both variables (Lagrangian coordinate and time). The linearization of this PDE leads to the well-known Eddington's equation of the standard model.

The case of no energy exchange between the polytrope and the outer medium is considered, that is, polytrope's motions are adiabatic. If gamma (a ratio of the specific heats of the gas) is 4/3 than PDE obtained allows the separation of variables. There exist two types of solutions of the problem both describing limitless expansion without shock wave formation. The first one is an expansion with positive total energy, and the second one is an expansion with zero total energy.

The second solution is of an astrophysical interest. It describes the permanently retarding expansion that, perhaps, is akin to a born of a red giant. The stellar density in this case concentrates to the centre of the star stronger than the density of the stationary star with the same gamma.

## INTRODUCTION

A self-gravitating gaseous sphere is one of the main objects of astrophysical study. Thermonuclear-burning stars, white and brown dwarfs can be modelled in such a manner. These objects can have extremely complex dynamics. They can be subjected to regular or chaotic pulsations, mass loss from their surfaces, limitless expansion or catastrophic compression. At present the main method to analyze these phenomena is a numerical modelling. Unfortunately, numerical results are often found to be obscure; because of this, it is highly desirable to amplify these results by analytical or semianalytical study of simple special cases. But already at the beginning of this way researchers came up against such great difficulties that this analysis is, in fact, in its infancy until the present time. In particular, even for the simplest case – one-dimensional (radial) motions of a polytrope – only stationary solutions and small oscillations near them are analytically studied. For both cases the problem reduces to ODE. These ODE's have no exact solutions and have to be integrated numerically.

In the present study the author sets himself the task of broadening the knowledge of radial motions of polytropes. Below is deducted PDE describing such motions in Lagrangian variables. It is shown that this equation allows the separation of variables if $\gamma = 4/3$. ODE's obtained are numerically integrated by the use of the analytical expansions near singular points.

---
[*] Russian Academy of Sciences, A. Ishlinsky Institute for Problems in Mechanics, Moscow, Russia;
*e-mail*: m-i-ivanov@mail.ru

# 1. FORMULATION OF THE PROBLEM

Write the basic equations of the problem in Lagrangian variables. Let a star has a radius $a$ and a density $\rho_0$ at an initial time $t = 0$. Denoting the radius, the density, and the pressure at an arbitrary time $t$ by $r$, $\rho$, and $p$, respectively, we write the system of equations:

$$\left(\frac{\partial^2 r}{\partial t^2} + \frac{Gm}{r^2}\right)\frac{\partial r}{\partial a} + \frac{1}{\rho}\frac{\partial p}{\partial a} = 0 \tag{1.1}$$

$$\frac{\partial m}{\partial a} = 4\pi r^2 \rho \frac{\partial r}{\partial a} \tag{1.2}$$

$$r^2 \rho \frac{\partial r}{\partial a} = a^2 \rho_0 \tag{1.3}$$

$$p = K\rho^\gamma \tag{1.4}$$

$$p = \frac{R_g}{\mu}\rho T \tag{1.5}$$

Here, (1.1) is the equation of motion, (1.2) is the equation for the gravity force, (1.3) is the continuity equation, (1.4) is the polytropic equation of state, and (1.5) is the equation of ideal gas; $m(a)$ is the stellar mass bounded by the sphere having the radius $a$ at the time $t = 0$, $T$ is a temperature, $G$ is the gravitational constant, $\gamma > 1$ is a ratio of the specific heats of the gas, $R_g$ is the universal gas constant, $\mu$ is an average molecular mass of gas, $K$ is some constant. The equation (1.5) just allows calculating a temperature and is not in use in the subsequent discussion.

Note that from the equations (1.2)-(1.3) it follows that the mass function is time-constant in Lagrangian variables and correlates with Lagrangian coordinate $a$ by one-to-one correspondence. Physically, it is equivalent of the fact that a sequence of stellar layers does not change at all radial motions. Thus we can use $m$ as a new variable in the system (1.1)-(1.4). Then, the system (1.1)-(1.4) reduces:

$$\left(\frac{\partial^2 r}{\partial t^2} + \frac{Gm}{r^2}\right)\frac{\partial r}{\partial m} + \gamma K\rho^{\gamma-2}\frac{\partial \rho}{\partial m} = 0 \tag{1.6}$$

$$4\pi r^2 \rho \frac{\partial r}{\partial m} = 1 \tag{1.7}$$

Enter a typical mass $M$ (a mass of the star), a typical radius $R$ (a radius of the star) and a typical time $\Omega = \sqrt{R^3/(\gamma GM)}$, and turn to dimensionless variables. Expressing the density from the equation (1.7) and substituting it in (1.6), we obtain the only equation for the dimensionless function $r(m,t)$:

$$krr_{mm} + 2kr_m^2 - \gamma^{-1}mr^{2\gamma-3}r_m^{\gamma+1} - r^{2\gamma-1}r_m^{\gamma+1}r_{tt} = 0 \tag{1.8}$$

Here, $k = (4\pi)^{1-\gamma} KG^{-1}R^{4-3\gamma}M^{\gamma-2}$ is the only dimensionless number of the problem.

The equation (1.8) has to be complemented by two boundary conditions and one initial condition. First, the polytrope's central density must be finite:

$$\lim_{m\to 0}\rho(m,t) < \infty$$

This can be expressed in dimensionless variables as:

$$\lim_{m\to 0} r(m,t)^{-2} r_m(m,t)^{-1} < \infty \tag{1.9}$$

Second, the polytrope's surface pressure must be equal to the pressure of an interstellar medium that we suppose to be zero:

$$\lim_{m \to M} p(m,t) = 0$$

This can be expressed in dimensionless variables as:

$$\lim_{m \to 1} r(m,t)^{-2} r_m(m,t)^{-1} = 0 \tag{1.10}$$

The initial condition is:

$$r(1,0) = 1 \tag{1.11}$$

that is, the star of the mass $M$ has the radius $R$ at the initial time $t = 0$.

The equation (1.8) is very difficult in the study. First, it is PDE of the second order in both variables. Second, it is nonlinear. In addition, this nonlinearity looks nothing like hydrodynamic one (for example, of Korteweg – de Vries or Burgers types), because of this, the known mathematical methods to analyze equations with hydrodynamic nonlinearities had failed in our case. Besides, analysis shows that the equation (1.8) does not possess the Painlevé property, and one cannot select the transformation that transform (1.8) to the equation possessing this property. Third, if $\gamma$ is non-integer (namely, the values of this sort are typical for real gases) than the equation (1.8) includes terms in which not only the function $r(m,t)$ but its first derivative enters in a fractional degree as well. Fourth, the equation (1.8) has singular coefficients since both the stellar centre $m=0$ and the stellar surface $m=1$ are its singular points.

## 2. TOTAL ENERGY OF THE STAR

Important information about the problem is provided by total energy of the star. It includes gravitational (potential), kinetic and internal terms $E = E_{\text{grav}} + E_{\text{kin}} + E_{\text{int}}$. In physical variables these terms are expressed as:

$$E_{\text{grav}} = -G \int_0^M \frac{m\,dm}{r} \tag{2.1}$$

$$E_{\text{kin}} = \frac{1}{2} \int_0^M v^2 dm \tag{2.2}$$

$$E_{\text{int}} = \frac{1}{\gamma-1} \int_0^M \frac{p}{\rho} dm \tag{2.3}$$

Turn in (2.1)-(2.3) to dimensionless variables:

$$E_{\text{grav}} = -E_0 \int_0^1 \frac{m\,dm}{r} \tag{2.4}$$

$$E_{\text{kin}} = E_0 \cdot \frac{\gamma}{2} \int_0^1 r_t^2 dm \tag{2.5}$$

$$E_{\text{int}} = E_0 \cdot \frac{k}{\gamma-1} \int_0^1 \left(r^2 r_m\right)^{1-\gamma} dm \tag{2.6}$$

where $E_0 = GM^2/R$ is typical energy.

Rearrange the term (2.6). From (1.8) we have:

$$\left(r^2 r_m\right)^{1-\gamma} = \frac{1}{2k\gamma}\frac{m}{r} + \frac{1}{2k} r r_{tt} - \frac{1}{2} r^{3-2\gamma} r_m^{-1-\gamma} r_{mm} \tag{2.7}$$

On the other hand, the following identity is valid:

$$r^{3-2\gamma} r_m^{-1-\gamma} r_{mm} = \frac{3-2\gamma}{\gamma} \left(r^2 r_m\right)^{1-\gamma} - \frac{1}{\gamma} \frac{\partial}{\partial m}\left(r^{3-2\gamma} r_m^{-\gamma}\right) \tag{2.8}$$

Substituting (2.8) in (2.7), we obtain:
$$\left(r^2 r_m\right)^{1-\gamma} = \frac{1}{3k}\frac{m}{r} + \frac{\gamma}{3k}rr_{tt} + \frac{1}{3}\frac{\partial}{\partial m}\left(r^{3-2\gamma}r_m^{-\gamma}\right) \qquad (2.9)$$

From (2.4)-(2.6) in view of (2.9) we obtain the formula for dimensionless total energy of the star in the form:
$$\frac{E}{E_0} = \frac{1}{3(\gamma-1)}\left(\int_0^1\left[(4-3\gamma)\frac{m}{r} + \gamma r r_{tt} + \frac{3}{2}\gamma(\gamma-1)r_t^2\right]dm + kr^{3-2\gamma}r_m^{-\gamma}\Big|_0^1\right) \qquad (2.10)$$

This formula ignores the boundary conditions (1.9)-(1.10). To take them into the consideration one need construct analytical series for the initial boundary value problem (1.8)-(1.11) in neighbourhoods of the singular points $m=0$ and $m=1$. Calculations show that these series are:
$$r(m,t) = Am^{1/3} + \frac{3^{1-\gamma}}{10k\gamma}A^{3(\gamma-1)}\left(\gamma A^2 A'' + 1\right)m + O\left(m^{5/3}\right),\ m\to 0,\ A=A(t) \qquad (2.11)$$

$$r(m,t) = B - \frac{\gamma}{\gamma-1}k^{\frac{1}{\gamma}}B^{2\left(\frac{2}{\gamma}-1\right)}\left(\gamma B^2 B'' + 1\right)^{-\frac{1}{\gamma}}(1-m)^{1-\frac{1}{\gamma}} + ...,\ m\to 1,\ B=B(t) \qquad (2.12)$$

where $A(t)$ and $B(t)$ are some unknown time functions. From the initial condition (1.11) it follows that $B(0)=1$.

From (2.11)-(2.12) it is easy to obtain that the term $kr^{3-2\gamma}r_m^{-\gamma}\Big|_0^1$ is identically zero. Hence, the formula (2.10) for the adiabatic case is:
$$\frac{E}{E_0} = \frac{1}{3(\gamma-1)}\int_0^1\left[(4-3\gamma)\frac{m}{r} + \gamma r r_{tt} + \frac{3}{2}\gamma(\gamma-1)r_t^2\right]dm \qquad (2.13)$$

## 3. EQUILIBRIUM (STATIONARY STATE) AND SMALL OSCILLATIONS NEAR IT

Stationary solution $r_0(m)$ describes the internal structure of a fully-convective star; the equation (1.8) becomes ODE, the corresponding boundary value problem has the form:
$$kr_0 r_0'' + 2kr_0'^2 - \gamma^{-1}mr_0^{2\gamma-3}r_0'^{\gamma+1} = 0 \qquad (3.1)$$
$$\lim_{m\to 0} m^{-1/3}r_0(m) = \text{const} \qquad (3.2)$$
$$\lim_{m\to 1} r_0'(m) = \infty \qquad (3.3)$$
$$r_0(1) = 1 \qquad (3.4)$$

The condition (3.2) is a direct consequence of (1.9). The condition (3.4) is a consequence of the initial condition (1.11) in the stationary case. The boundary value problem is not overdeterminate since the dimensionless number $k$ is unknown. The solution of the problem (3.1)-(3.4) is well known [1].

The formula (2.13) for the problem (3.1)-(3.4) takes the form:
$$\frac{E}{E_0} = \frac{4-3\gamma}{3(\gamma-1)}\int_0^1 \frac{m}{r_0}dm \qquad (3.5)$$

Hence it follows that total energy of the star is negative ($E<0$) for $\gamma > 4/3$, because of this, such star is stable. If $1<\gamma<4/3$ than total energy is positive ($E>0$) and the star is unstable. For $\gamma = 4/3$ total energy is zero and the star is in indifferent equilibrium [2].

Consider small oscillations near equilibrium $r_0(m)$. We will search for the solution in the form:

$$r(m,t) = r_0(m) + y(m)\exp(i\sigma t) \tag{3.6}$$

where the function $y(m)$ is small in comparison with $r_0(m)$. All the quantities entering in the formula (3.6) are dimensionless, and $\sigma$ is a dimensionless eigenfrequency. Substitute (3.6) in (1.8), keep only the terms of the first order of smallness, and express $r_0"$ from (3.1). Then:

$$kr_0^2 y" + \left(4kr_0 r_0' - \frac{\gamma+1}{\gamma} mr_0^{2(\gamma-1)} r_0'^\gamma \right) y' + \left(\sigma^2 r_0^{2\gamma} r_0'^{\gamma+1} - 2kr_0'^2 + 2\frac{2-\gamma}{\gamma} mr_0^{2\gamma-3} r_0'^{\gamma+1}\right) y = 0 \tag{3.7}$$

The equation obtained turns to the well-known Eddington's equation in going to the function $y/r_0$ [3].

Obtain the boundary conditions for (3.7). Substitute the ansatz (3.6) in the boundary conditions (1.9) and (1.10), take into account that $r_0(m)$ satisfies the boundary conditions (3.2)-(3.4), and neglect terms of second and higher orders of smallness. Hence, we obtain the boundary conditions for the function $y(m)$:

$$y(0) = 0 \tag{3.8}$$
$$y'(1) + 2r_0'(1) y(1) = 0 \tag{3.9}$$

The boundary value problem (3.7)-(3.9) is the so-called standard model that is well known from the literature [3].

The standard model spectrum is discrete and all the eigenfrequencies $\sigma$ are real. To the smallest eigenvalue corresponds the function $y(m)$ without zeros – the so-called fundamental tone of stellar radial pulsations. As one goes to the next eigenvalue number the number of zeros of the function $y(m)$ always increases by one.

## 4. THE FIRST SPECIAL CASE OF ADIABATIC MOTION

Let $\gamma = 4/3$. Then the equation (1.8) transforms to:

$$krr_{mm} + 2kr_m^2 - \frac{3m}{4} r^{-1/3} r_m^{7/3} - r^{5/3} r_m^{7/3} r_{tt} = 0 \tag{4.1}$$

where $k = (4\pi)^{-1/3} KG^{-1} M^{-2/3}$, that is, does not depend on the stellar radius $R$.

One can see that the equation (4.1) is satisfied by:
$$r(m,t) = H(m)(t+t_0) \tag{4.2}$$

where $t_0 > 0$ is some constant, and the function $G(m)$ obeys the boundary value problem:

$$4kH^{4/3} H" - 3mH'^{7/3} + 8kH^{1/3} H'^2 = 0 \tag{4.3}$$
$$\lim_{m \to 0} m^{-1/3} H(m) = \text{const} \tag{4.4}$$
$$\lim_{m \to 1} H'(m) = \infty \tag{4.5}$$

In the neighbourhood of the singular point $m = 0$ the function $H(m)$ is represented as:
$$H(m) = H_1 m^{1/3} + O(m), \quad m \to 0 \tag{4.6}$$

where $H_1$ is some constant.

Integration of the problem (4.3)-(4.5) presents difficulties because of the condition (4.5) is singular. In order to circumvent this situation we will use the function $H$ as a new variable and the variable $m$ as a new function. In addition, denominate $Z = H(1)$ and $h = H/Z$. With $H' = Z(dm/dh)^{-1}$ and $H" = -Z(d^2m/dh^2)(dm/dh)^{-3}$ we transform the equation (4.3) to the form:

$$4kh^{4/3}\frac{d^2m}{dh^2}+3m\left(\frac{dm}{dh}\right)^{2/3}-8kh^{1/3}\frac{dm}{dh}=0 \qquad (4.7)$$

The condition (4.4) transforms to the form:
$$m(0)=0 \qquad (4.8)$$
and the condition (4.5) transforms to two boundary conditions:
$$m(1)=1 \qquad (4.9)$$
$$m'(1)=0 \qquad (4.10)$$

The additional condition (4.9) comes into being since we enter the unknown parameter $Z$ into the problem. This parameter is wanted to determine. In this special case $Z$ does not appear in the equation (4.7). Since the series for $m(h)$ near $h=0$ has the form:
$$m(h)=(Z/H_1)^3 h^3 + O(h^5), \; h\to 0 \qquad (4.11)$$
that a solution of the problem (4.7)-(4.10) will depend only on $Z/H_1$ and $k$.

Using the expansion (4.11), one can integrate the problem (4.7)-(4.10) by the method of the local separation of singularities that has been proposed by the author in [4]. Previously, this method has been applied only for integrations of linear ODE's [4-8]. But, as numerical calculations show, it can be easily extended to nonlinear cases. In particular, the method, as applied to the problem (4.7)-(4.10), finds solutions that are stable with respect to both including of higher terms in the series (4.11) and varying of the radius of the singular point neighbourhood; these facts are the main criteria of plausibility of the results obtained.

The integration shows that there exists the unique solution of the problem (4.7)-(4.10) – $k=0.15654$ and $Z/H_1=3.75$. Hence, the equation (4.3) has the continuum of solutions in this case. It is well known that there exists the continuum of stellar equilibriums for $\gamma=4/3$ [2], and the equation (4.3) is equivalent of the equation that describes the stationary state of such stars; because of this, the result obtained makes sense. The initial condition (1.11) will be satisfied if:
$$t_0=Z^{-1} \qquad (4.12)$$

From (1.7) we find the stellar density in the form:
$$\rho=\frac{1}{4\pi Z^3 h^2}\frac{dm}{dh}(t+t_0)^{-3} \qquad (4.13)$$
With (4.12) we obtain:
$$\rho=\frac{1}{4\pi h^2}\frac{dm}{dh}(Zt+1)^{-3} \qquad (4.14)$$

Find total stellar energy for the solution (4.2). From (2.13) we obtain:
$$E_G=\frac{2E_0}{3}\int_0^1 H^2 dm >0 \qquad (4.15)$$

This result is easily understood. Since the equation (4.3) is equivalent of the equation describing the stationary state, the sum of the potential and internal terms is identical to that for the stationary problem, that is, zero, as obtained above. But, in our case the star in addition expands, because of this, it must have nonzero kinetic term. Thus total energy is always positive in this case.

## 5. THE SECOND SPECIAL CASE OF ADIABATIC MOTION

Let consider the second case of adiabatic motion. The equation (4.1) is also satisfied by:
$$r(m,t)=F(m)(t+t_0)^{2/3} \qquad (5.1)$$
where $t_0>0$ is some constant, and the function $F(m)$ obeys the boundary value problem:

$$36kF^{4/3}F''+\left(8F^3-27m\right)F'^{7/3}+72kF^{1/3}F'^2=0 \tag{5.2}$$

$$\lim_{m\to 0} m^{-1/3}F(m)=\text{const} \tag{5.3}$$

$$\lim_{m\to 1} F'(m)=\infty \tag{5.4}$$

In the neighbourhood of the singular point $m=0$ the function $F(m)$ is represented as:

$$F(m)=F_1 m^{1/3}+O(m),\ m\to 0 \tag{5.5}$$

where $F_1$ is some constant.

Again, we will use the function $F$ as a new variable and the variable $m$ as a new function. In addition, denominate $Z=F(1)$ and $f=F/Z$. Then, the equation (5.2) transforms to the form:

$$36kf^{4/3}\frac{d^2m}{df^2}-\left(8Z^3 f^3-27m\right)\left(\frac{dm}{df}\right)^{2/3}-72kf^{1/3}\frac{dm}{df}=0 \tag{5.6}$$

The condition (5.3) transforms to the form:

$$m(0)=0 \tag{5.7}$$

and the condition (5.4) transforms to two boundary conditions:

$$m(1)=1 \tag{5.8}$$

$$m'(1)=0 \tag{5.9}$$

Analogously, the additional condition (5.8) comes into being since we enter the unknown parameter $Z$ into the problem. From (5.5) we obtain the series for $m(f)$ near $f=0$ in the form:

$$m(f)=(Z/F_1)^3 f^3+O(f^5),\ f\to 0 \tag{5.10}$$

Similarly, using the expansion (5.10), one can integrate the problem (5.6)-(5.9) by the method of the local separation of singularities. In this way we find three parameters of the problem – $k$, $F_1$ and $Z$. The parameter $t_0$ may be found from the condition (1.11).

The integration shows that, unlike the previous case, there exist a group of solutions in the narrow interval of values of $k$ between $0.1519$ and $0.1531$. The initial condition (1.11) will be satisfied if:

$$t_0=Z^{-3/2} \tag{5.11}$$

The solution parameters are shown in the table.

| $k$ | $F_1$ | $Z$ | $t_0$ |
|---|---|---|---|
| 0.1519 | 0.28207 | 1.49 | 0.550 |
| 0.1525 | 0.27063 | 1.24 | 0.724 |
| 0.1531 | 0.25763 | 1.13 | 0.832 |

From (1.7) we find the stellar density in the form:

$$\rho=\frac{1}{4\pi Z^3 f^2}\frac{dm}{df}(t+t_0)^{-2} \tag{5.12}$$

With (5.11) we obtain:

$$\rho=\frac{1}{4\pi f^2}\frac{dm}{df}\left(Z^{3/2}t+1\right)^{-2} \tag{5.13}$$

Find total stellar energy for the solution (5.1). From (2.13) we obtain:

$$E_F=0 \tag{5.14}$$

Thus the star expanding in this way has zero total energy as the stationary star with the same $\gamma$.

Illustrate the results obtained. Some radius-density dependencies are shown in the Fig. 1. The density is shown in the logarithmic scale, and $t=0$, that is, we consider the initial moment of the expansion. There exists only one radius-density function for the solution (4.2); the dependency on $Z$ is absent. This function is portrayed by the solid line. Other curves show the solutions of the form (5.1). The density for $k=0.1519$ is portrayed by the dashed line, and the density for $k=0.1531$ is portrayed by the dash-and-dot line. One can see that the stellar matter for the second-type solution (5.1) concentrates to the centre of the star stronger than one for the first-type solution (4.2). Since the radius-density function of the first-type solution coincides with the stationary one it may be deduced that the strong concentration of the stellar matter to the centre of the star in the second case produces more strong gas pressure so that even zero total energy is sufficiently that the star begins expanding.

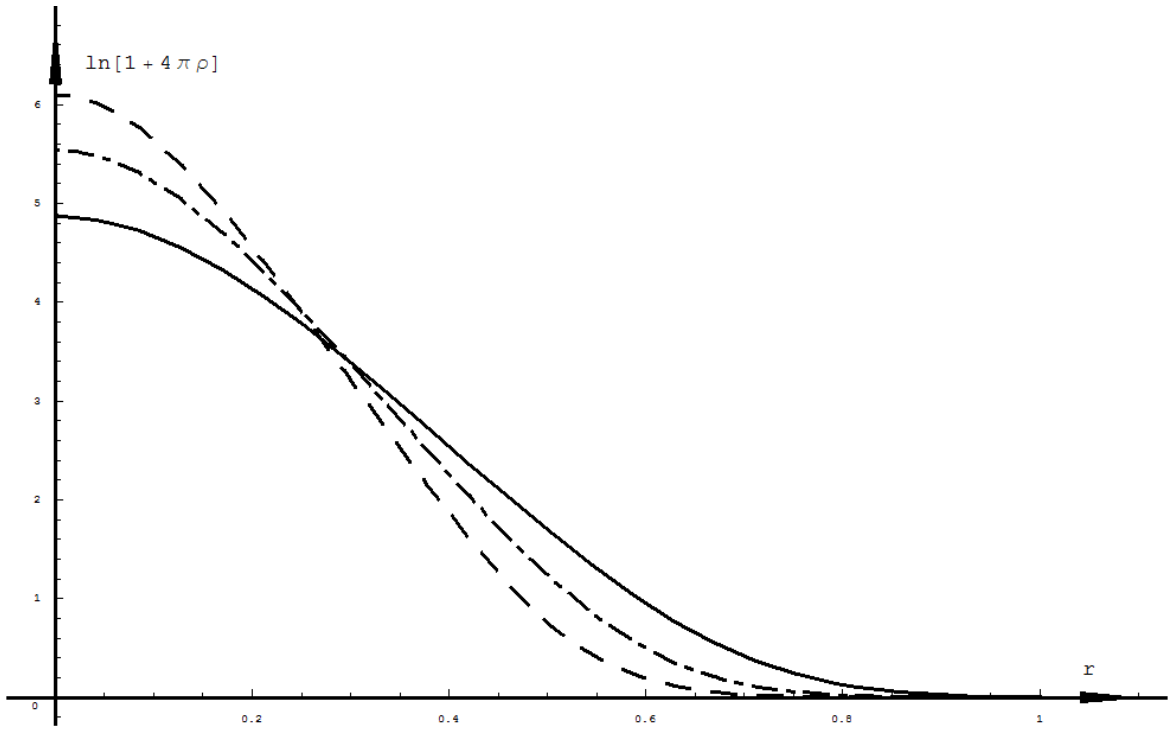

**Fig. 1.**

The radius-density dependencies at the time $t=5$ are shown in the Fig. 2. Since the dependency of the solution (4.2) on $Z$ is not absent for $t \neq 0$ that this solution is portrayed by two solid lines – for $Z=1.13$ (the density in the stellar centre is higher) and for $Z=1.49$ (the density in the stellar centre is lower). These results are clear – bigger values of $Z$ correspond to the solution with bigger total energy $E_G$, and the star expands speeder in this case. Other curves show the solutions of the form (5.1). The density for $k=0.1519$ is portrayed by the dashed line, and the density for $k=0.1531$ is portrayed by the dash-and-dot line. On can see that the expansion is speeder in the first case ($k=0.1519$) because of bigger value of $Z$, and within some time interval the central density of the second case ($k=0.1531$) becomes higher than one in the first case. As before, concentration to the stellar centre is stronger for the second-type solutions; dissimilarity between the solutions of different types comes into particular prominence.

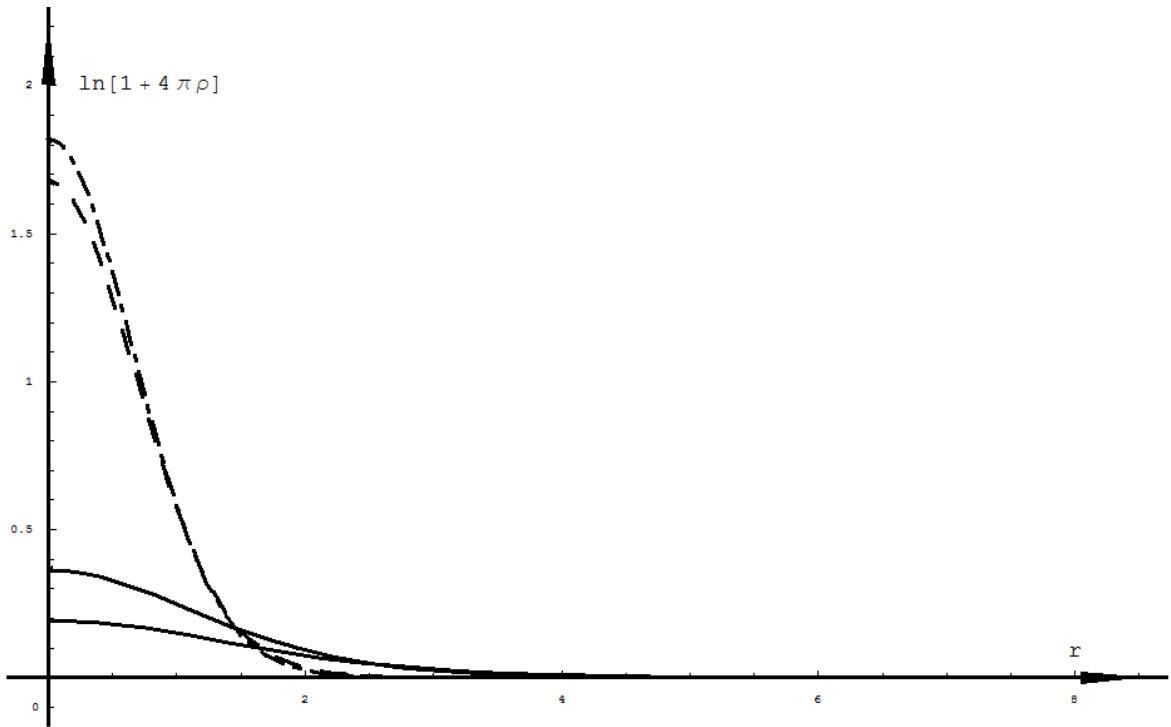

**Fig. 2.**

## 6. DISCUSSION

The motions obtained are both limitless expansions without shock wave formation. This property does them to look like the Einstein-de Sitter expansion of spatially-isothermal gas [9]. But, in our case the gaseous sphere expands adiabatically that is correlated to the special structure of the sphere expanded.

It is common knowledge that a stable polytropic star has $\gamma > 4/3$ and negative total energy [2]. During stellar evolution $\gamma$ decreases and total energy increases reaching zero at $\gamma = 4/3$. In this moment the star stops to be stable. Then it either restores negative total energy by a burst or destructs.

Since the solution (4.2) has nonzero positive total energy, in order to a star expands in this manner it must be cross an area of instability without destructing that seems to be incredible. In the case of the solution (5.1) we have a completely different situation. Since the expansion in this case occurs with zero total energy, that is, in indifferent equilibrium, it can readily imagine following evolution of a polytropic star: until $\gamma$ remains more than $4/3$ the star is stable, but as soon as the critical value is attained the expansion in the form (5.1) begins. It is common knowledge that stars near the critical value of $\gamma$ oscillates with very long periods, for example, mirids [10]. It is not improbable that the process (5.1) is the limit case of such oscillation with the infinitely long period.

One detail is noteworthy. Usually a loss of stellar stability is accompanied by the core collapse and shock wave formation in the expanding stellar envelope [11]. In our case this is not so. The expansion is smooth and any contraction does not occur. This is somewhat strange situation, and in order to elucidate the nature of the process found we consider a concrete example. Let take a star like the Sun with the radius $R = R_\odot = 7 \cdot 10^8$ m and the mass $M = M_\odot = 2 \cdot 10^{30}$ kg. Calculate: in what time will this star expand to $2R$? The result is: from 23 to 35 minutes depending on the value of $k$. This time is quite small and the process can be reasonably called an explosion. But it is a retarding expansion. Thus let calculate: in what time will our star expand to the Earth's orbit with the radius $1.5 \cdot 10^{11}$ m? The result is: from 4.5 to 6.8 years depending on the value of $k$. There is a very slow expansion to be called an explosion. Note that more massive stars have lower densities and, consequently, its expansions will still slower. Hence, this process is best-borne similarity to a born of a red giant. This assumption is supported by the fact that, observing long-period variables, astronomers see only motions of the stellar matter away from the star and never to it [12], that is, such stars continue expanding.

This work is supported by RFBR (project 11-01-00247).